\documentclass[10pt]{article}
\usepackage{amssymb,latexsym, amsmath}
\usepackage{amscd}

\newtheorem{theorem}{Theorem}

\newcommand{\R}{\mathbb{R}}

\DeclareMathOperator{\ann}{\mathrm{ann}}
\newcommand{\ad}{\mathrm{ad}}
\DeclareMathOperator{\Ad}{\mathrm{Ad}}

 \DeclareMathOperator{\pr}{\mathrm{pr}}
\begin{document}

\title{Magnetic Geodesic Flows on Coadjoint Orbits
\footnote{{\bf MSC:} 70H06, 37J35, 53D25}
\footnote{Journal Ref: J. Phys. A: Math. Gen. {\bf 39} (2006) L247 - L252 }
\footnote{Doi:10.1088/0305-4470/39/16/L01}}

\author{Alexey V. Bolsinov
 , \,
Bo\v zidar Jovanovi\' c  \\
\\
\small Department of Mechanics and Mathematics, Moscow State University\\
\small 119992, Moscow, Russia, e-mail: bolsinov$@$mech.math.msu.su \\
\small and\\
\small Mathematical Institute SANU\\
\small Kneza Mihaila 35, 11000 Belgrade, Serbia, e-mail: bozaj$@$mi.sanu.ac.yu}

%\date{}
\maketitle

\begin{abstract}
We describe a class of completely
integrable $G$-invariant magnetic geodesic flows on (co)adjoint
orbits of a compact connected Lie group $G$ with magnetic field given by the
Kirillov-Konstant 2-form. 
\end{abstract}

\section{Introduction}

Let $Q$ be a smooth manifold with a
local coordinate system $x^1,\dots,x^n$ and Riemannian metric $g=(g_{ij})$.
The inertial motion of the unit mass particle under the influence of
the additional magnetic field given by a closed
2-form
$$
\Omega=\sum_{1 \le i<j \le n} F_{ij}(x) dx^i \wedge dx^j,
$$
is described by the following equations on the phase space $T^*Q$:
\begin{equation}
\frac{dx^i}{dt}=\frac{\partial H}{\partial p_i},
\qquad \frac{dp_i}{dt}=-\frac{\partial H}{\partial x^i}+
\sum_{j=1}^{n} F_{ij} \frac{\partial H}{\partial p_j},
\label{magnetic_flow}
\end{equation}
where $p_j= g_{ij}\dot x^i$ are canonical momenta and
the Hamiltonian is
$
H(x,p)=\frac{1}{2}\sum g^{ij}p_i p_j.
$
Here $g^{ij}$ are the coefficients of the tensor inverse to the metric.

The equations (\ref{magnetic_flow}) are Hamiltonian with
respect to the symplectic form $\omega+\rho^*\Omega$,
where $\omega=\sum dp_i\wedge dx^i$ is the canonical symplectic form on $T^*Q$ and
$\rho: T^*Q\to Q$ is the natural projection.
Namely, the corresponding Poisson bracket is given by
\begin{equation}
\{f,g\}=\sum_{i=1}^{n}\left( \frac{\partial f}{\partial x^i}\frac{\partial g}{\partial p_i}
-\frac{\partial g}{\partial x^i}\frac{\partial f}{\partial p_i}\right)+
\sum_{i,j=1}^{n} F_{ij} \frac{\partial f}{\partial p_i}\frac{\partial g}{\partial p_j},
\label{magnetic_bracket}
\end{equation}
and the Hamiltonian equations $\dot f=\{f,H\}$ read (\ref{magnetic_flow}).
The flow (\ref{magnetic_flow}) is called {\it magnetic geodesic flow} on the Riemannian manifold
$(Q,g)$ with respect to the magnetic field $\Omega$.

In this paper we consider $G$-invariant magnetic geodesic flows on
(co)adjoint orbits $\mathcal O$ of a compact connected Lie group
$G$, where  $\Omega$ is the Kirillov-Konstant 2-form (Theorem \ref{th1}).
The non-commutative integrability of the systems, for the normal
metrics, is proved recently by Efimov \cite{Ef1, Ef2}. Following
\cite{BJ2}, we  give a new, short proof of the non-commutative
integrability (Theorem \ref{th2}). In addition, the usual Liouville integrability by
means of commuting analytic integrals is shown. One can use the
commuting integrals in order to deform the normal metric to a
certain class of $G$-invariant metrics on $\mathcal O$ with
completely integrable magnetic geodesic flows as well (Theorem \ref{commutative}).

\section{Magnetic Coadjoint Orbits}

Let $G$ be a compact connected Lie group with the Lie algebra $\mathfrak g=T_e G$.
Let us fix some $\Ad_G$-invariant scalar product $\langle\cdot,\cdot\rangle$ on $\mathfrak g$.
By the use of $\langle\cdot,\cdot\rangle$ we identify $\mathfrak g^*$ with $\mathfrak g$.

Consider the adjoint action of $G$ and the $G$-orbit $\mathcal O(a)=\Ad_G(a)$
through an element  $a\in\mathfrak g$.
Let $\xi\in\mathfrak g$ and $x=\Ad_g(a)$. Since
\begin{equation}
\xi_x=\frac{d}{ds}\Ad_{\exp(s\xi)}(x)\vert_{s=0}=[\xi,x],
\label{xi}
\end{equation}
the tangent space $T_x\mathcal O(a)$ is simply $[\mathfrak g,x]$, i.e.,
it is the orthogonal complement to
$\ann(x)=\{\eta\in\mathfrak g\,\vert\,[\eta,x]=0\}$.
By definition, the Kirillov-Konstant symplectic form $\Omega$ on
$\mathcal O(a)$ is a $G$-invariant form, given by
\begin{equation}
\Omega(\eta_1,\eta_2)\vert_{x}=
-\langle x, [\xi_1,\xi_2]\rangle, \quad   \eta_i=[\xi_i,x], \quad i=1,2.
\label{Kirillov}
\end{equation}
Similarly, the scalar product $\langle \cdot,\cdot\rangle$
induces the {\it normal metric} $K_0$
on $\mathcal O(a)$ as follows
\begin{equation}
K_0(\eta_1,\eta_2)\vert_{x}=
\langle \xi_1,\xi_2\rangle, \quad   \eta_i=[\xi_i,x], \quad i=1,2.
\label{K_0}
\end{equation}

The cotangent bundle $T^*\mathcal O(a)$ can be realised as a submanifold of
$\mathfrak g\times \mathfrak g$
$$
T^*\mathcal O(a)=\{(x,p)\, \vert\, x=\Ad_g(a), p\in\ann(x)^\perp\},
$$
with the paring between $p\in T^*_x \mathcal O(a)$
and $\eta\in T_x\mathcal O(a)$ given by $p(\eta)=\langle p,\eta\rangle$.
Then the canonical symplectic form $\omega$ on $T^*\mathcal O(a)$
can be seen as a restriction of the canonical linear symplectic
form of the ambient space $\mathfrak g\times\mathfrak g$:
$
\sum_{i=1}^{\dim\mathfrak g} dp_i \wedge dx_i,
$
where $p_i$, $x_i$
are coordinates of $p$ and $x$ with respect to some base of $\mathfrak g$.

The  $G$-action
\begin{equation}
g\cdot (x,p)=(\Ad_g x,\Ad_g p)
\label{action}
\end{equation}
is Hamiltonian on $(T^*\mathcal O(a),\omega)$. From (\ref{xi})
we find that
the momentum mapping is given by the relation
$
\langle \Phi_0(x,p),\xi \rangle=\langle p,\xi_x\rangle=\langle p,[\xi,x]\rangle.
$
That is
$$
\Phi_0(x,p)=[x,p].
$$

Following Efimov \cite{Ef1, Ef2}, we consider magnetic geodesic flows on $\mathcal
O(a)$ with respect to the magnetic fields $\epsilon\Omega$, where
$\Omega$ is Kirillov-Konstant 2-form (\ref{Kirillov}) and
$\epsilon\in\R$. According to (\ref{magnetic_flow}), the adding of
magnetic field $\epsilon\Omega$ to the system reflects as a
perturbation of the system in $p$-variable by the magnetic force
$\Pi_\epsilon$, determined by $\langle
\Pi_\epsilon,\eta\rangle=-\epsilon\langle x,
[\ad_x^{-1}\eta,\ad_x^{-1}\dot x]\rangle$, $\eta\in T_x\mathcal
O(a)$. Hence
$
\Pi_\epsilon=-\epsilon \ad_x^{-1}\dot x.
$

The $G$-action (\ref{action}) is Hamiltonian on $(T^*\mathcal
O(a),\omega+\epsilon\Omega)$ as well \cite{Ef2, MP}. In our
notation we have that the momentum mapping reads
$$
\Phi_\epsilon(x,p)=\Phi_0(x,p)+\epsilon x=[x,p]+\epsilon x.
$$

\paragraph{G-Invariant Magnetic Geodesic Flows.}
The $G$-invariant metrics on $\mathcal O(a)$ are in one-to-one correspondence
with $\Ad_{G_a}$-invariant positive definite operators
$$
\varphi: \mathfrak v \to \mathfrak v, \quad \Ad_g\circ \varphi=\varphi\circ \Ad_g, \quad g\in G_a,
$$
where
$\mathfrak v=T_a\mathcal O(a)=\ann(a)^\perp$ and $G_a$ is the isotropy
group of $a$.
Namely, for a given $\varphi$, we define
$$
\varphi_x=\Ad_g\circ\varphi\circ \Ad_{g^{-1}}: 
T_x\mathcal O(a)\to T_x\mathcal O(a), \quad x=\Ad_g(a),$$
and a $G$-invariant metric
$K_\varphi(\eta_1,\eta_2)\vert_x=\langle \varphi_x \eta_1,\eta_2\rangle$. After Legendre
transformation $T\mathcal O(a)\to T^*\mathcal O(a)$ with respect to $K_\varphi$,
we get the Hamiltonian function for the given metric:
$$
H_\varphi(x,p)=\frac12\langle \varphi^{-1}_x p,p\rangle.
$$

\begin{theorem}\label{th1}
The equations of the magnetic geodesic flow on $(\mathcal O(a),K_\varphi)$
with respect to the magnetic term $\epsilon\Omega$, in redundant
variables $(x,p)$, are given by
\begin{eqnarray}
&& \dot x=\varphi^{-1}_x p, \label{x} \\
&& \dot p=\ad_x^{-1}[p,\varphi^{-1}_x p]-\pr_{\ann(x)}[\ad_x^{-1}\varphi_x^{-1}p,p]-\epsilon\ad_x^{-1}\varphi_x^{-1} p.\label{p}
\end{eqnarray}
In particular, the magnetic flow of the normal metric (\ref{K_0}) reads
\begin{eqnarray}
&& \dot x=[[x,p],x], \label{xx} \\
&& \dot p=[[x,p],p] +  \epsilon [x,p],\label{pp}
\end{eqnarray}
\end{theorem}

\noindent{\it Proof.}
The equation (\ref{x}) is just the inverse of the Legendre transformation.
We can derive (\ref{p}) simply by using the conservation of the momentum
mapping $\Phi_\epsilon$ for $G$-invariant Hamiltonians.
We have
\begin{eqnarray}
\frac{d}{dt}\Phi_\epsilon(x,p)&=&[\dot x,p]+[x,\dot p]+\epsilon \dot x=0 \nonumber\\
&=& [\varphi^{-1}_x p,p]+[x,\dot p]+\epsilon[x,\ad_x^{-1}\varphi^{-1}_x p]=0.
\label{dot-Phi}
\end{eqnarray}
Since $\varphi^{-1}$ is $\Ad_{G_a}$-invariant, the term $[\varphi^{-1}_x p,p]$
belongs to $\ann(x)^\perp$.
Thus from (\ref{dot-Phi}) we get
\begin{equation}
\pr_{\ann(x)^\perp} \dot p=
\ad_x^{-1}[p,\varphi^{-1}_x p]-\epsilon\ad_x^{-1}\varphi_x^{-1} p.
\label{p1}
\end{equation}

In order to find $\pr_{\ann(x)}\dot p$, take the (local)
orthonormal base $e_1(x),\dots,e_r(x)$ of  $\ann(x)$.
Then $\pr_{\ann(x)}\dot p$ is determined from the condition that
the trajectory $(x(t),p(t))$ satisfies constraints
\begin{equation}
\frac d{dt}\langle p,e_i(x)\rangle = \langle \dot
p,e_i(x)\rangle+\langle p,\dot e_i(x)\rangle=0, \quad i=1,\dots,r.
\label{lambda}
\end{equation}

From $[e_i(x),x]\equiv 0$, $i=1,\dots,r$,
we get
\begin{equation}
[\dot e_i(x),x]+[e_i(x),\dot x]=[\dot e_i(x),x]+
[e_i(x),[x,\ad_x^{-1}\varphi^{-1}_x p]]=0
\quad i=1,\dots,r. \label{e_i}
\end{equation}

Furthermore, combining (\ref{e_i}) and  the
Jacobi identities
$$
[e_i,[x,\ad_x^{-1}\varphi^{-1}_x p]]+[x, [\ad_x^{-1}\varphi^{-1}_x p,e_i]]+
[\ad_x^{-1}\varphi^{-1}_x p, [e_i,x]]=0,\quad i=1,\dots ,r
$$
we obtain $\dot e_i(x)=[e_i(x),\ad_x^{-1}\varphi^{-1}_x p]$
(modulo $\ann(x)$). Whence, using (\ref{lambda}) we get $\langle
\dot p,e_i(x)\rangle+\langle [\ad_x^{-1}\varphi^{-1}_x
p,p],e_i\rangle=0$, $i=1,\dots,r$, i.e,
\begin{equation}
\pr_{\ann(x)}\dot p=-\pr_{\ann(x)}[\ad_x^{-1}\varphi^{-1}_x p,p].
\label{p2}
\end{equation}
The relations (\ref{p1}) and (\ref{p2}) proves (\ref{p}). 

Now, for
the normal metric $K_0$ we have
$\varphi_x=-\ad^{-1}_x\circ\ad^{-1}_x$ and the Hamiltonian is
\begin{equation}
H_0=-\frac12 \langle \ad_x\ad_x p,p\rangle=\frac12 \langle [x,p],[x,p]\rangle=
\frac12\langle \Phi_0(x,p),\Phi_0(x,p)\rangle.
\label{H}
\end{equation}
The equation (\ref{xx}) follows directly from (\ref{x}), while (\ref{p1}) and (\ref{p2}) become
\begin{eqnarray*}
&&\pr_{\ann(x)^\perp} \dot p=
\ad_x^{-1}[p,[x,[p,x]]]+\epsilon[x,p], \\
&&\pr_{\ann(x)}\dot p=\pr_{\ann(x)}[[x, p],p].
\end{eqnarray*}
Again, the Jacobi identity gives
$$
[p,[x,[p,x]]=[x,[[x,p],p]]=\ad_x(\pr_{\ann(x)^\perp}[[x,p],p])
$$
which together with the above formulae proves (\ref{pp}). $\Box$

\medskip

The geometry of the Hamiltonian flows on cotangent bundles,
in this representation, is studied by Bloch, Brockett and Crouch \cite{BBC}.
The system (\ref{xx}), (\ref{pp}), for $\epsilon=0$, agrees with the equations (2.7) given in \cite{BBC},
while the system (\ref{x}), (\ref{p}) differs from 
the equations (2.19) \cite{BBC}.
The equations (2.19) \cite{BBC} describe the geodesic flows of
submersion (or collective) metrics on the orbit $\mathcal O(a)$, and, in general,
are not $G$-invariant. Recall that the submersion metrics are given
by Hamiltonians of the form $H=\frac12\langle\Phi_0(x,p),\phi\,\Phi_0(x,p)\rangle$,
where $\phi$ is a symmetric, positive definite operator on $\mathfrak g$.
Specially, $K_0$ is both $G$-invariant and  submersion metric.

\section{Integrable Flows}

Let $\mathcal F_1^\epsilon$ be the algebra of all analytic, polynomial in momenta, 
functions of the form
$\mathcal F_1^\epsilon=\{p\circ\Phi_\epsilon \,\vert\, p\in \R[\mathfrak g]\}$
and $\mathcal F_2$ be the algebra of all analytic, polynomial in
momenta, $G$-invariant functions on $T^*\mathcal O(a)$. Then, according to
the Noether theorem
$$
\{\mathcal F_1^\epsilon,\mathcal F_2\}_\epsilon=0,
$$
where
$\{\cdot,\cdot,\}_\epsilon$ are magnetic Poisson bracket with
respect to
$
\omega+\epsilon\rho^*\Omega.
$

Consider the Hamiltonian $H_\epsilon=\frac12 \langle
\Phi_\epsilon,\Phi_\epsilon \rangle\in \mathcal F_1^\epsilon$.
A simple calculation shows $H_\epsilon(x,p)=H_0+\epsilon^2\frac12 \langle a,a \rangle$.
Thus, we see that Hamiltonian flows of $H_0$ and
$H_\epsilon$ coincides. Since $H_{\epsilon}$ belongs
to $\mathcal F_1^\epsilon$ its commutes with $\mathcal F_2$.
On the other side, as a composition of the momentum mapping with an invariant polynomial, the
function $H_\epsilon$ is also $G$-invariant and commutes with $\mathcal F_1^\epsilon$.
From the above consideration and Theorem 2.1 \cite{BJ2} we 
recover the Efimov result \cite{Ef2}:

\begin{theorem}\label{th2}
Let $G$ be a compact Lie group and $a\in \mathfrak g$. The 
magnetic geodesic flows of normal metric (\ref{xx}), (\ref{pp}) on the 
adjoint orbit $\mathcal O(a)$ is completely integrable in the non-commutative sense. 
\end{theorem} 

Namely, the algebra of first integrals
$
\mathcal F_1^\epsilon+\mathcal F_2
$
is complete on $(T^*\mathcal O(a),\omega+\epsilon\rho^*\Omega)$ (see \cite{BJ2}) and its
invariant level sets are isotropic tori. Similarly as in the Liouville theorem, the tori
are filled up with quasi-periodic 
trajectories of the system (\ref{xx}), (\ref{pp}) (see \cite{MF2, N}).

\paragraph{Integrable Deformations.}
Let $\mathcal A\subset \R(\mathfrak g)$ be a commutative set
of polynomials with respect to Lie-Poisson brackets on $\mathfrak g$. One can
always find $\mathcal A$ that is complete on generic orbits
$\mathcal O(\Phi_\epsilon(x,p))$ (e.g, see \cite{Bo}). Let
$\Phi^*_\epsilon{\mathcal A}$ be the pull-back of $\mathcal A$ by the momentum map: 
$\Phi^*_\epsilon\mathcal A=\{h\circ \Phi_\epsilon\, \vert\, h\in \mathcal A\}$.

Let $\mathcal B$ be a commutative subset of $\mathcal{F}_2$,
with respect to the magnetic Poisson bracket.
Then
$\Phi^*_\epsilon{\mathcal A}+{\mathcal B}$ is a complete commutative set on
$(T^*\mathcal O(a),\omega+\epsilon\rho^*\Omega)$ if
$\mathcal B$ is a complete commutative subset of
$\mathcal F_2$, i.e., we have
\begin{equation}
\delta=\dim \mathcal O(a) - \frac12\dim \mathcal O(\Phi_\epsilon(x,p))
\label{delta}\end{equation}
independent functions in $\mathcal B$,
for a generic element $(x,p)\in T^*\mathcal O(a)$ \cite{BJ3}.

The $G$-invariant, polynomial in momenta functions $f(x,p)$ on
$T^*\mathcal O(a)$, are in one-to-one  correspondence with
$\Ad_{G_a}$-invariant polynomials  on $\mathfrak v$ via
restriction to $T^*_a\mathcal O(a)$: $f_0(p_0)=f(a,p_0)$. Next, we
apply the transformation 
$$
f_0\mapsto \bar f, \quad f_0=\bar f\circ \Phi_0\vert_{x=a}=\bar f\circ \ad_a.
$$
Within these identifications, from (\ref{magnetic_bracket}),
(\ref{Kirillov}) and Thimm's formula for $\epsilon=0$ \cite{Th},
the magnetic Poisson bracket $\{f,g\}_\epsilon(x,p)$ corresponds
to the following bracket (our notation is slightly different from Efimov's \cite{Ef2}) 
\begin{equation}
\{\bar f(\mu),\bar g(\mu)\}^\epsilon_\mathfrak v=-\langle \mu+\epsilon a,
[\nabla \bar f(\mu),\nabla \bar g(\mu) ]\rangle,  \label{b2}
\end{equation}
where $\mu=[a,p_0]$, $x=\Ad_g a$, $p=\Ad_g p_0$.

Note that
$\{\{\cdot,\cdot\}^\lambda_\mathfrak v, \, \lambda\in\mathbb{R}\}$
is a pencil of the compatible Poisson
brackets on the algebra of $\Ad_{G_a}$-invariant polynomials
$\mathbb{R}[\mathfrak v]^{G_a}$.
By the use of this pencil and the completeness
criterion derived in \cite{Bo}, it is proved
that the family of polynomials
\begin{equation}
{\mathcal B}_a=\{p^\lambda_a(\mu)=p(\mu+\lambda a), \;
\lambda\in\mathbb{R},\;p\in \mathbb{R}[\mathfrak g]^G,\; \eta\in
\mathfrak v \}
\label{shift}
\end{equation} is a complete
commutative subset of $\mathbb{R}[\mathfrak v]^{G_a}$ with respect
to the canonical brackets $\{\cdot,\cdot\}^0_\mathfrak v$ (see \cite{BJ3, MP}).
Here $\mathbb{R}[\mathfrak g]^G$ is the algebra of
$\Ad_{G}$-invariant polynomials on $\mathfrak g$. 
Using the method of \cite{Bo}, it can be verified
that $\mathcal B_a$ is a complete commutative set with
respect to the magnetic Poisson bracket (\ref{b2}) as well.

Let $b$ an element from the center of $\ann(a)$.
Define the {\it sectional operator} $\bar\phi_{a,b}: \mathfrak v\to \mathfrak v$ by
$\bar\phi_{a,b}=\ad_a^{-1}\circ \ad_b=\ad_b\circ\ad_a^{-1}$.
For compact groups, among sectional operators we can take positive definite ones.
It easily follows from \cite{MF1} that
the function  $\bar H_{a,b}=\frac12\langle \bar\phi_{a,b}(\mu),\mu\rangle$
belongs to $\mathcal B_a$.
The corresponding $G$-invariant function is
$$
H_{a,b}(x,p)=\frac12\langle \ad_{b_x} p,\ad_x p
\rangle=-\frac12\langle \ad_x\ad_{b_x}p,p\rangle=\frac12 \langle \phi_{x,b_x} p,p\rangle,
$$
where $b_x=\Ad_g b$, $x=\Ad_g a$ and $\phi_{x,b_x}=-\ad_x\ad_{b_x}$.
(Recall that $b$ belongs to the center of $\ann(a)$ and
since $G$ is compact connected Lie group, $G_a$ is also connected,
so $b_x$ is well defined.)
This is a Hamiltonian function 
of the $G$-invariant metric $K_{a,b}$:
\begin{equation}
K_{a,b}(\eta_1,\eta_2)=
\langle (\ad_{b_x})^{-1}\eta_1,\ad_x^{-1}\eta_2\rangle, 
\label{K_ab}
\end{equation}
where $\eta_1,\eta_2\in T_x\mathcal O(a)$. Whence, we get the following statement

\begin{theorem} \label{commutative}
The magnetic geodesic flows of the metrics $K_{a,b}$ with respect to the
magnetic term $\epsilon\Omega$:
\begin{eqnarray*}
&& \dot x=-\ad_x\ad_{b_x} p=[[b_x,p],x], \\
&& \dot p=-\ad_x^{-1}[p,[x,[b_x,p]]]+\pr_{\ann(x)}[[b_x,p],p]+\epsilon [b_x,p]
\end{eqnarray*}
are completely integrable
in the commutative sense, by means of analytic, polynomial in
momenta first integrals.
\end{theorem}

The Liouville Lagrangian tori are additionally foliated by
$\delta$-dimensional invariant isotropic tori, level sets
of integrals $\mathcal F_1^\epsilon+\mathcal B_a$ 
($\delta$ is given by (\ref{delta})).
Note that $\delta$ does not depend on $\epsilon$:
for a generic $\eta \in\mathfrak v$ we have equality
$\dim \mathcal O(\eta+\epsilon a)=\dim \mathcal O(\eta)$
for all $\epsilon \in \mathbb{R}$ (see \cite{BJ3, MP}).
Therefore, the influence of the magnetic fields $\epsilon\Omega$,
$\epsilon \in \mathbb{R}$ reflects as a deformation of the
foliation of the phase space
$T^*\mathcal O(a)$ by invariant tori. As the magnetic field increases, the 
magnetic geodesic lines become more curved.

\paragraph{Concluding Remarks.}
One can take $b$ such that the operator $\phi_{x,b_x}$ is positive, but with kernel
different from zero. Then the Hamiltonian flow of $H_{a,b}$, for $\epsilon=0$, represents
an integrable 
{\it sub-Riemannian} geodesic flow on the orbit $\mathcal O(a)$ with the constraint
distribution $D$ at the point $x$ given by the image 
$$
D_x=\phi_{x,b_x}(T_x^*\mathcal O(a))=\ad_{b_x} (\ann(x))^\perp\subset T_x\mathcal O(a)
$$
and the sub-Rieamannian structure defined by (\ref{K_ab}),
where now $\eta_1,\eta_2\in D_x$. Here we assume that the 
distribution $D$ is bracket generating (see \cite{M, BJ3} for more details).

There is a natural generalization of the above results to the
class of magnetic potential systems on coadjoint orbits as well as to
the wider class of homogeneous spaces. We shall consider these
problems in the forthcoming paper.

\paragraph{Acknowledgments.}
The first author was supported by Russian Found for Basic Research, RFBR 05-01-00978. The second author
was supported by the Serbian Ministry of Science, Project "Geometry and Topology of Manifolds and Integrable Dynamical Systems".

\end{document}